\documentclass[a4paper,twocolumn,11pt]{article}
\usepackage[spanish]{babel}
\usepackage{ae}
\usepackage[latin1]{inputenc} 
\usepackage[T1]{fontenc} 
\usepackage{layout}
\usepackage{array}
\usepackage{graphicx}
\usepackage{amsmath}
\usepackage{multicol}
\hyphenrules{nohyphenation} 
\usepackage{rotating} 
\usepackage{multirow} 
\usepackage[]{cite}   
\usepackage{anysize}  
\usepackage{endnotes} 

\usepackage{tikz,fp,ifthen,fullpage}
\usetikzlibrary{arrows,snakes,backgrounds}
\usepackage{pgfplots}
\usetikzlibrary{shapes,trees}
\usetikzlibrary{calc,through,backgrounds,decorations}
\usetikzlibrary{decorations.shapes,decorations.text,decorations.pathmorphing,backgrounds,fit}
\usetikzlibrary{fadings,intersections}
\usetikzlibrary{patterns}
\usetikzlibrary{mindmap}
\marginsize{1.5cm}{1.0cm}{1.5cm}{2cm}

\begin{document}

\renewcommand{\tablename}{Tabla}
\renewcommand{\abstractname}{}
\renewcommand{\thefootnote}{\arabic{footnote}}

\title{
        {\LARGE Estimación del tiempo de iluminación solar sobre la tierra mediante un modelo analítico: un escenario fértil para enseñar física}\\       
\footnotesize   \textit{(Estimation of solar illumination time on the earth by an analytical model: a fertile scenery for to teach physics)}
}
 
\author
{ 
Paco Talero $^{1,3}$, Fernanda Santana $^{2}$\\
César Mora$^{3}$,\\
$^{1}$ {\small Grupo F\'{\i}sica y Matem\'{a}tica, Dpt de Ciencias
Naturales, Universidad Central},\\ 
\small {Carrera 5 No 21-38, Bogot\'{a}, D.C. Colombia.},\\ 
$^{2}$ {\small Observatorio Astronómico Nacional de Colombia}\\
\small {Universidad Nacional De Colombia,Carrera 45 No 26-85, Bogotá D.C. Colombia}\\
$^{3}$ {\small Centro de Investigaci\'{o}n en Ciencia Aplicada y Tecnolog\'{\i}a Avanzada del Instituto Polit\'{e}cnico Nacional,} \\
\small {Av. Legaria 694, Col. Irrigaci\'{o}n, C. P. 11500, M\'{e}xico D. F.}
}

\date{}
\twocolumn
[
\begin{@twocolumnfalse}
\maketitle 
\begin{abstract}
Se formuló un modelo analítico que permitió  estimar el tiempo de iluminación solar sobre la Tierra para cualquier fecha del año y cualquier latitud, el modelo tomó  la oblicuidad de la eclíptica constante, los rayos de luz  paralelos, la Tierra esférica y el movimiento de la Tierra circular uniforme, también mostró un contexto de la astronomía para enseñar física básica. Se relacionó  el movimiento de la Tierra alrededor del Sol con  el movimiento del plano de  luz proyectado sobre la Tierra, luego se dedujo  la zona iluminada para una latitud dada y se calculó el tiempo de iluminación mediante el movimiento circular uniforme de rotación terrestre. El modelo se confrontó con resultados numéricos de la Geoscience Australia Agency  hallando como error porcentual máximo $1,6\%$, el cual se atribuyó principalmente a la discrepancia entre la trayectoria real elíptica y la trayectoria circular tomada en este modelo. Así, sin hacer uso de la trigonometría esférica se obtuvo un modelo analítico que  explica de manera muy aproximada el tiempo de iluminación solar en cualquier época del año y para cualquier latitud, el modelo brinda un contexto auténtico  para estudiar algunos aspectos de la física básica.\\
\textbf{Palavras-chave:} modelo, astronomía, enseñanza de la física, métodos de enseñanza.\\ 

We proposed an analytical model for the calculus of illumination time of the Earth for any time of year and any latitude, this model assumes the obliquity of the ecliptic as constant, the light beams as parallels, the Earth as spherical, the movement of translation of Earth as uniform circular, also this model showed a context of the astronomy whereby the teachers can teach the basic physics.It was built through a relationship between the movement of translation and of rotation of the  wave front light,  then we  found the of  illumination zone  on the Earth  and  the illumination time is estimated  in a particular latitude with the uniform circular movement of Earth.  Present model was confronted with the numerical results of the Geoscience Australia Agency and it is found a maxim perceptual error of $1,6\%$, this value was assigned primarily to the difference between the circular trajectory, in this model, and the elliptical trajectory that is the real. Without the use of spherical trigonometry was obtained an analytical model that estimates very close  the  solar illumination time at any time of year and  any latitude on earth, the model provides an authentic context for studying basic aspects of physics.

\end{abstract}
\end{@twocolumnfalse}
]

\section{Introducción}

Recientes investigaciones en enseñanza de la física han venido mostrando que la astronomía trae consigo contextos auténticos de alta  motivación en los estudiantes que han permitido desarrollar diversos contenidos físicos a diferentes niveles de formación, tales contextos abarcan tópicos como la Tierra, la escala del sistema solar, la determinación de distancias a los planetas y a las estrellas, la ley de Hubble, la ley de Wien y las curvas de rotación en galaxias espirales, entre otros \cite{Edi,Planet,Trayec,SimuF,EXV,Santa,CurvaR,Exec,Luna,Hubble,Wien}. Así mismo,  estos contextos han permitido  introducir contenidos físicos tales como dinámica, cinemática y óptica. En \cite{Planet} se muestra la experiencia de buscar planetas extrasolares a través de la adquisición de datos propios mediante el control remoto de un telescopio diseñado para tal fin; en \cite{Trayec} se expone un estudio que revela algunas ideas erróneas sobre la trayectoria de los planetas alrededor del sol; en \cite{SimuF,EXV} se estudia   el movimiento planetario a través de experimentos virtuales con base en argumentos físicos elementales e intuitivos que permiten desarrollar discusión sobre la ley de gravitación, la tercera ley de Kepler y métodos numéricos de implementación simple; en \cite{CurvaR} se estudian curvas de rotación en galaxias espirales ;en \cite{Exec} se muestra como obtener  la excentricidad de la órbita terrestre usando, principalmente,  las leyes de Kepler y un instrumento de observación sencillo; en \cite{Luna} se repasan  algunos  métodos geométricos mediante los cuales se estima  las distancias  Tierra-Luna y Tierra-Sol, así como  los diámetros del Sol y la Luna en relación al radio de la Tierra; en \cite{Hubble}  se aprovecha el contexto de la ley de Hubble para promover la habilidad de la interpretación conceptual en gráficas de  velocidad contra distancia y en \cite{Wien} se muestra como una presentación desde  la estadística de fotones de la ley de radiación de cuerpo negro es más eficaz, desde el punto de vista pedagógico, que el tratamiento tradicional que se hace generalmente en los libros de texto.   

Dentro del contexto anterior,  este trabajo muestra como el fenómeno cotidiano del día y la noche puede usarse para estudiar aspectos básicos de física como factores de conversión,  movimiento circular uniforme (MCU) y  óptica de rayos. El problema concreto de estudio consiste en responder la pregunta: ¿es posible estimar mediante un modelo analítico basado en conceptos físicos elementales el tiempo de iluminación solar  sobre la Tierra para cualquier fecha del año y en cualquier latitud?  Para responder esta pregunta se desarrolla un modelo analítico que deja de lado algunos hechos astrofísicos que no son relevantes durante el transcurso de pocos años, así las características del modelo son la siguientes:  toma en cuenta la corrección estándar de la refracción de la luz por la atmósfera y el tiempo diario de iluminación debido al movimiento de traslación de la Tierra alrededor del Sol;  desprecia los efectos tanto de precesión  como de  nutación;  considera la Tierra completamente esférica; asume  el movimiento de traslación como un MCU y toma los rayos de luz solar paralelos. 

Para la formulación del modelo se establece el sistema de referencia inercial con  uno de sus ejes paralelo al eje de rotación de la Tierra (considerado estático respecto a las estrellas fijas) y los demás ejes también anclados a estrellas fijas. Ahora, se establece un vínculo entre el MCU de la Tierra alrededor del Sol  y el movimiento del frente de onda plano de luz solar el cual  gira sobre la esfera terrestre, luego se delimita la zona de la Tierra iluminada en cualquier fecha del año a una latitud arbitraria y se aplican conceptos de cinemática del MCU de rotación de la Tierra para calcular el tiempo que un punto terrestre estaría iluminado. Mediante las ecuaciones obtenidas es posible conocer de manera muy aproximada el tiempo de iluminación para una fecha particular del año en todo el mundo, conocer la fecha de ocurrencia del Sol de media noche para cualquier latitud y mostrar diagramas de simetría útiles  a la hora de interpretar hechos físicos mediante la lectura de gráficos.

El modelo permite realizar análisis gráfico del tiempo de iluminación durante un año para una latitud determinada, estas gráficas permiten realizar una confrontación sencilla del modelo con resultados numéricos obtenidos de simulaciones que toman en cuenta los efectos de otros fenómenos astrofísicos en particular de la trayectoria elíptica de la Tierra, al confrontar el modelo con los resultados arrojados en el simulador de la Geoscience Australia Agency  se halla un error porcentual máximo en la gráfica mencionada muy cercano a $1,6\%$ error que se atribuye  fundamentalmente, y de acuerdo con la tercera ley de Kepler, a la diferencia entre las trayectorias elíptica y circular. 

Este trabajo muestra, sin hacer uso de la trigonometría esférica,  una manera de pensar físicamente sobre el hecho cotidiano de la noche y el día, a través de un modelo analítico basado fundamentalmente en MCU que a través de sus resultados permite explicar de manera muy aproximada el tiempo de iluminación solar en cualquier época del año y para cualquier latitud, el modelo deja un escenario cautivante dentro de la astronomía para estudiar algunos aspectos básicos de la física como el MCU y la óptica de  rayos paralelos.  
 
Este artículo está organizado de manera siguiente: en la sección ($2$) se deduce el tiempo de iluminación terrestre de acuerdo con  lo planteado anteriormente,  en la sección ($3$)  se hacen correcciones por traslación y refracción, en la sección ($4$) se comparan los resultados del modelo con  resultados numéricos, en la sección ($5$) se proponen algunas estrategias didácticas para poner en práctica los resultados de este modelo y en la sección ($6$) se muestran las conclusiones.

\section{Tiempo de iluminación solar sobre la Tierra}  

Se sabe que debido a la rotación de la Tierra sobre su propio eje esta  no posee una geometría que corresponda por completo a una esfera sino más bien a un elipsoide \cite{Karttunen,Green,Portilla}.Debido a la geometría elipsoidal de la Tierra y a la interacción gravitacional con la Luna, el Sol y los demás cuerpo celestes que componen el sistema solar el movimiento de traslación y rotación terrestre varia respecto a lo que se espera del estudio de un sistema Sol- Tierra completamente aislado. Sin embargo, estos efectos no se toman en cuenta en este trabajo debido a que  sus efectos son despreciables durante el transcurso de un año\cite{Green,Alonso}.   

A lo anterior se añade que la luz solar se toma con  la aproximación de rayos paralelos, esto a causa de que  observaciones bien establecidas  muestran una desviación del paralelismo de aproximadamente  $0,5^{o}$,que es producida por la refracción de la luz al atravesar la atmósfera\cite{Paul}.
\begin{figure}[ht]
\begin{center}
\begin{tikzpicture}[xscale=1.0,yscale=1.0]
\draw[thick,black] (0cm,0cm) circle (4cm); 
\draw[dashed,black] (-4.5,0)--(4.5,0); 
\draw[dashed,black] (0,-4.5)--(0,4.0);
\draw[dashed,black] (-2.2,4.4)--(2.2,-4.2);
\draw[dashed,black,line width=1pt] (-3.12,2.5)--(3.12,2.5);
\draw[dashed,black,line width=1pt] (3.12,2.5)--(3.12,0.0);
\draw[thick,black,line width=1pt] (0cm,0cm)--(3.12,2.5);
\draw[-latex,gray,line width=8pt] (-4cm,-1.82) -- (-1cm,-0.455cm);
\draw[-latex,line width=0.5pt] (0cm,1cm) arc (90:113.5:1cm);
\draw[-latex,line width=1pt] (1cm,0cm) arc (0:36:1cm);
\fill[black]  (-1.23cm,2.5cm) circle(0.08cm); 
\fill[black]  (0cm,2.5cm) circle(0.08cm); 
\fill[black]  (3.12cm,2.5cm) circle(0.08cm); 
\fill[black]  (0cm,0cm) circle(0.08cm); 
\coordinate [label=below:\textcolor{black} {PNT}] (x) at  (0cm,4.5cm);
\coordinate [label=below:\textcolor{black} {$\epsilon$}] (x) at  (-0.25cm,1.5cm);
\coordinate [label=below:\textcolor{black} {$\phi$}] (x) at  (1.3cm,0.7cm);
\coordinate [label=below:\textcolor{black} {$p$}] (x) at  (-1.1cm,3.05cm);
\coordinate [label=below:\textcolor{black} {$o$}] (x) at  (-0.2cm,-0.01cm);
\coordinate [label=below:\textcolor{black} {$q$}] (p) at  (0.2cm,3.0cm);
\coordinate [label=below:\textcolor{black} {$s$}] (x) at  (3.3cm,2.9cm);
\coordinate [label=below:\textcolor{black} {$M$}] (x) at  (2.5cm,-3.7cm);
\coordinate [label=below:\textcolor{black} {$L$}] (x) at  (0.2cm,-4.1cm);
\coordinate [label=below:\textcolor{black} {$h$}] (x) at  (3.3cm,1.3cm);
\coordinate [label=below:\textcolor{black} {$R$}] (x) at  (1.5cm,1.8cm);
\coordinate [label=below:\textcolor{black} {Luz solar}] (x) at  (-2cm,-1.5cm);
\coordinate [label=below:\textcolor{black} {Ecuador}] (x) at  (2.8cm,-0.1cm);
\end{tikzpicture}
\caption{Corte meridional de la Tierra paralelo a los rayos de luz provenientes del Sol  para  el solsticio de invierno en el hemisferio norte.}
\label{SolsI}\end{center}
\end{figure}
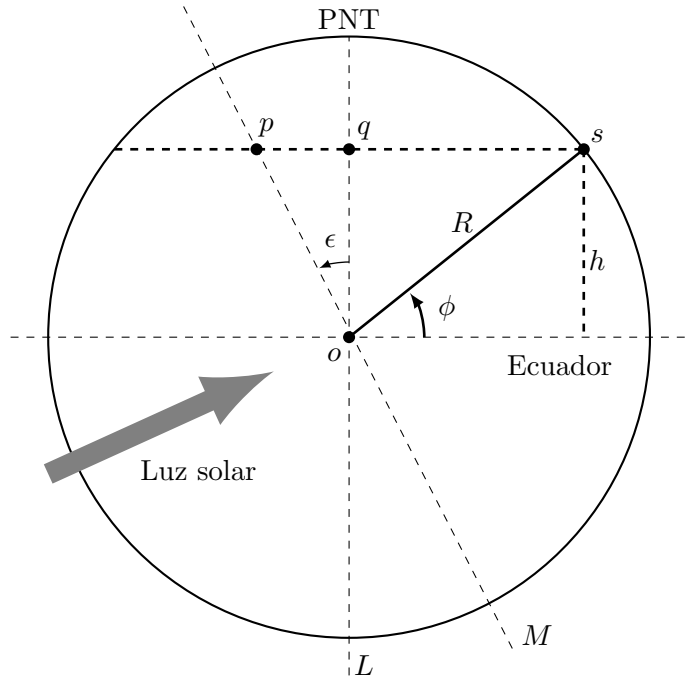

En la Fig.\ref{SolsI}  se muestra un corte meridional de la Tierra paralelo a los rayos de luz provenientes del Sol  para  el solsticio de invierno en el hemisferio norte,  se observa que el ángulo $\epsilon$ formado entre el plano $M$ del frente de onda de la luz solar y el eje de rotación de la Tierra $L$ es la oblicuidad de la eclíptica, que es el ángulo formado por el eje de rotación de la Tierra y el plano de traslación alrededor del Sol. Esto implica que un observador en la superficie de la Tierra a una latitud $\phi$ tendrá una trayectoria circular con radio $r =\overline{qs}$,  una velocidad angular $\omega$ y un periodo de rotación $T=\frac{2\pi}{\omega}$ que corresponden a la velocidad angular y al periodo de rotación terrestre como muestra la Fig.\ref{SolsI}.Así mismo, de la la Fig.\ref{SolsI} se puede obtener las ecuaciones
\begin{equation}\label{pch1}
\tan \phi=\frac{h}{r}
\end{equation}
donde $r_o=\overline{pq}$ y 
\begin{equation}\label{pch2}
\tan \epsilon=\frac{r_o}{h} .
\end{equation}
Al eliminar $h$ de (\ref{pch1}) y (\ref{pch2}) se obtiene
\begin{equation}\label{pch3}
\frac{r_o}{r}=\tan\phi\!\tan\epsilon.
\end{equation}
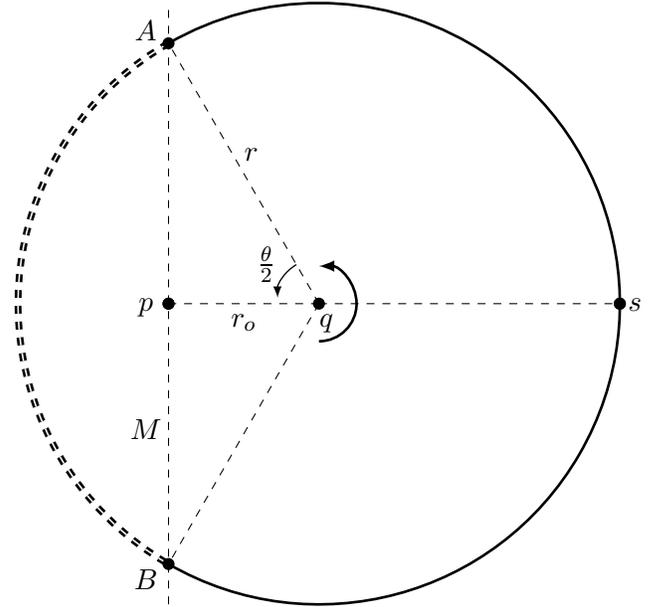
\begin{figure}[ht]
\begin{center}
\begin{tikzpicture}[xscale=1.0,yscale=1.0]
\draw[line width=1pt] (-2cm,-3.464cm) arc(240:480:4cm);
\draw[dashed,black,double,line width=1pt] (-2cm,3.464cm) arc(120:240:4cm);
\draw[dashed,black] (-2cm,-4cm)--(-2cm,4cm);
\draw[dashed,black] (0cm,0cm)--(-2cm,3.464cm);
\draw[dashed,black] (0cm,0cm)--(-2cm,-3.464cm);
\draw[dashed,black] (0cm,0cm)--(4cm,0cm);
\draw[dashed,black] (0cm,0cm)--(-2cm,0cm);
\fill[black] (-2cm,3.464cm) circle(0.08cm); 
\fill[black] (-2cm,-3.464cm) circle(0.08cm); 
\fill[black] (-2cm,0cm) circle(0.08cm); 
\fill[black] (4cm,0cm) circle(0.08cm); 
\fill[black] (-2cm,0cm) circle(0.08cm); 
\fill[black] (4cm,0cm) circle(0.08cm); 
\fill[black] (0cm,0cm) circle(0.08cm); 
\draw[-latex,line width=1pt] (0cm,-0.5cm) arc (270:450:0.5cm);
\draw[-latex,line width=0.5pt] (-0.3cm,0.5196cm) arc (120:181:0.5cm);
\coordinate [label=below:\textcolor{black} {$q$}] (x) at  (0.1cm,0.0cm);
\coordinate [label=below:\textcolor{black} {$s$}] (x) at  (4.2cm,0.2cm);
\coordinate [label=below:\textcolor{black} {$A$}] (x) at  (-2.3cm,3.9cm);
\coordinate [label=below:\textcolor{black} {$B$}] (x) at  (-2.3cm,-3.4cm);
\coordinate [label=below:\textcolor{black} {$M$}] (x) at  (-2.3cm,-1.4cm);
\coordinate [label=below:\textcolor{black} {$p$}] (x) at  (-2.3cm,0.2cm);
\coordinate [label=below:\textcolor{black} {$r_o$}] (x) at  (-1.0cm,0.0cm);
\coordinate [label=below:\textcolor{black} {$r$}] (x) at  (-0.9cm,2.2cm);
\coordinate [label=below:\textcolor{black} {$\frac{\theta}{2}$}] (x) at  (-0.7cm,0.9cm);
\end{tikzpicture}
\caption{Corte de la Tierra a latitud $\phi$ vista desde PNT.}
\label{SolsPNT}\end{center}
\end{figure}	
En la Fig.\ref{SolsPNT} se observa la Tierra aún en solsticio de invierno vista desde el polo norte terrestre (PNT), se observa  que la zona iluminada es el arco de circunferencia $\stackrel{\textstyle\frown}{\mathrm{AB}}$ con ángulo $\theta$ y radio $r=\overline{Aq}$. Además, la geometría evidenciada  que  el ángulo $\theta$ se puede expresar como $\cos\left(\frac{\theta}{2}\right)=\frac{r_o}{r}$, lo cual permite expresar con ayuda de  (\ref{pch3}) el ángulo $\theta$ a través de la ecuación
\begin{equation}\label{Ang}
 \cos\left(\frac{\theta}{2}\right)=\tan\phi\: \tan \epsilon.
\end{equation}
Ahora, si se supone que el plano $M$ no rota significativamente al rededor del punto $p$ durante una revolución terrestre el tiempo de iluminación $t_{D}$ en el solsticio se puede obtener al calcular el tiempo que un observador sobre la superficie terrestre tarda en barrer el ángulo $\theta$ con velocidad angular $\omega$, siendo $\omega=\frac{\theta}{t_D}$. Así se obtiene 
\begin{equation}\label{time1}
t_{D}=\frac{T}{\pi}\cos^{-1}\left(\tan\phi\: \tan\epsilon \right).
\end{equation}
El tiempo $t_{N}$ que  el observador permanece en la zona no iluminada se puede calcular a partir de (\ref{time1}) y es simplemente $t_{N}=T-t_{D}$.

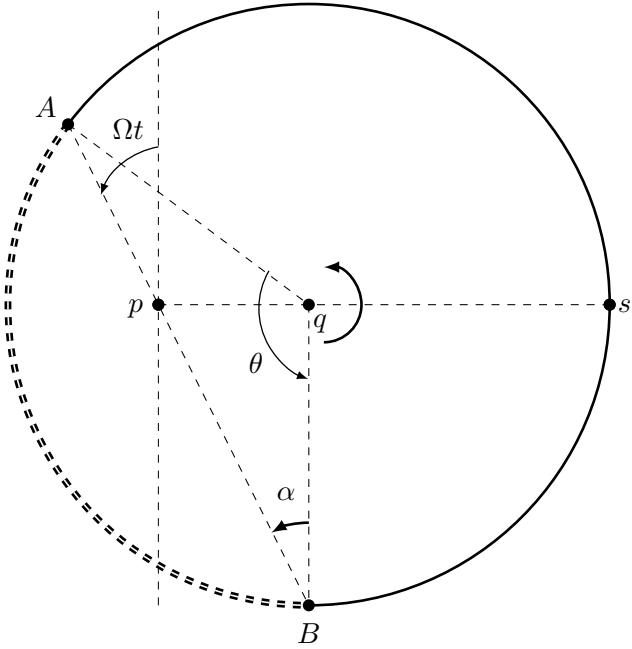
\begin{figure}[ht]
\begin{center}
\begin{tikzpicture}[xscale=1.0,yscale=1.0]
\draw[line width=1pt] (0cm,-4cm) arc (270:503.13:4cm);
\draw[-latex,line width=1pt] (0cm,-2.9cm) arc (90:120:1cm);
\draw[dashed,double,line width=1pt] (-3.2cm,2.4cm) arc (143.13:270:4cm);
\draw[dashed,black] (-3.2cm,2.4cm)--(0cm,-4cm);
\draw[dashed,black] (-3.2cm,2.4cm)--(0cm,0cm);
\draw[dashed,black] (0cm,0cm)--(0cm,-4cm);
\draw[dashed,black] (-2cm,0cm)--(4cm,0cm);
\draw[dashed,black,line width=0.1pt] (-2cm,-4cm)--(-2cm,4cm);
\fill[black] (-3.2cm,2.4cm) circle(0.08cm); 
\fill[black] (0cm,-4cm) circle(0.08cm); 
\fill[black] (-2cm,0cm) circle(0.08cm); 
\fill[black] (4cm,0cm) circle(0.08cm); 
\fill[black] (-2cm,0cm) circle(0.08cm); 
\fill[black] (4cm,0cm) circle(0.08cm); 
\fill[black] (0cm,0cm) circle(0.08cm); 
\draw[-latex,line width=1pt] (0.2cm,-0.5cm) arc (270:450:0.5cm);
\draw[-latex,line width=0.5pt] (-0.53cm,0.45cm) arc (150:250:1.0cm);
\draw[-latex,line width=0.5pt] (-2.0cm,2.1cm) arc (100:160:1.0cm);
\coordinate [label=below:\textcolor{black} {$q$}] (x) at  (0.15cm,0.0cm);
\coordinate [label=below:\textcolor{black} {$s$}] (x) at  (4.2cm,0.2cm);
\coordinate [label=below:\textcolor{black} {$A$}] (x) at  (-3.5cm,2.9cm);
\coordinate [label=below:\textcolor{black} {$B$}] (x) at  (0cm,-4.1cm);
\coordinate [label=below:\textcolor{black} {$p$}] (x) at  (-2.3cm,0.2cm);
\coordinate [label=below:\textcolor{black} {$\theta$}] (x) at  (-0.7cm,-0.5cm);
\coordinate [label=below:\textcolor{black} {$\alpha$}] (x) at  (-0.3cm,-2.3cm);
\coordinate [label=below:\textcolor{black} {$\Omega t$}] (x) at  (-2.4cm,2.6cm);
\end{tikzpicture}
\caption{Posición angular del plano $M$ en tiempo $t$.}
\label{Rotada}
\end{center}
\end{figure}

Al tomar como origen de tiempo el solsticio de invierno en el hemisferio norte la traslación de la Tierra alrededor del Sol se ve reflejada en la rotación del plano $M$ alrededor del punto $p$, ver Fig.\ref{Rotada}. Si se considera el movimiento de traslación como un MCU con velocidad angular $\Omega$ entonces el plano $M$ girará sobre $p$ barriendo un ángulo $\Omega t$ a partir de la posición inicial, como muestra la Fig.\ref{Rotada}.
Nótese en la Fig.\ref{Rotada} que el triángulo $\widehat{qAB}$ es isósceles y por lo tanto al usar la suma de ángulos internos se tiene $\theta=\pi-2\alpha$, al aplicar el teorema del seno al triángulo $\widehat{qqB}$ se encuentra  $r \sin\alpha=r_{o} \cos\left(\Omega t\right)$ y usar la ecuacion (\ref{pch3}) se encuentra
\begin{equation}\label{time2}
t_{D}=\frac{T}{\pi}\cos^{-1}\left(\tan\phi \: \tan\epsilon \cos\left(\Omega t\right) \right ).
\end{equation}
Si se considera la traslación de la Tierra alrededor del Sol como un MCU con $\Omega<<\omega$, se desprecia la rotación del plano $M$ del frente de onda de la luz solar y también se desprecia la refracción de la luz por la atmósfera la ecuación (\ref{time2}) permite estimar el tiempo de luz solar durante una revolución de la Tierra para cualquier fecha del año y cualquier latitud. 

La ecuación (\ref{time2}) tiene sentido sólo si se cumple la condición 
\begin{equation}\label{cond}
\left| \tan\phi \: \tan\epsilon \: \cos\left(\Omega t\right) \right| \leq 1,
\end{equation}
para que pueda evaluarse la función arcocoseno. Esta condición demarca la latitud máxima $\phi_{max}$ que puede ser evaluada en  (\ref{time2}) de acuerdo con el tiempo $t$ transcurrido. Así, de acuerdo con (\ref{cond}) se tiene para la máxima latitud la ecuación
\begin{equation}\label{Lmax}
\phi_{max}=\tan^{-1}\left(  \frac{\pm 1}{  \tan\epsilon \: \cos\left(\Omega t\right) } \right),
\end{equation}
que demarca el comienzo de Sol de media noche y la noche polar, esto sin tomar en cuenta el Leve incremento en el tiempo de iluminación debido a refracción de la luz y traslación de la Tierra. 
\section{Corrección de $t_D$ por traslación y refracción}
Durante una revolución de la Tierra sobre su eje el plano $M$ no permanece en reposo,  como se supuso, sino que rota un poco alrededor del punto $p$. En la Fig.\ref{RotaT} se muestra el ángulo de la zona iluminada inicial $\theta_i$, es decir justo cuando un observador sobre la superficie terrestre llega al punto $A$ y se muestra también un ángulo $\theta_f$ de la zona iluminada cuando el mismo observador justo alcanza el plano  $M$ por su otro extremo en el punto ${D}$. De acuerdo con esto el observador barre  un ángulo $\beta$ adicional, que es preciso calcular para determinar el tiempo adicional de iluminación. 

En la Fig.\ref{RotaT} se observa la relación entre los ángulos $\gamma$, $\alpha_{i}$ y $\Omega\tau$
\begin{equation}\label{R1}
\gamma+\alpha_{i}+ \Omega\tau=\pi,
\end{equation}
donde el tiempo $\tau$ se refiere al tiempo transcurrido desde que un observador ligado a la Tierra con latitud $\phi$ coincide con el plano de $M$ en el punto $A$ hasta que nuevamente coincide en el tiempo $t+\tau$, siendo $t$ el tiempo transcurrido desde el solsticio de invierno en el hemisferio norte hasta que observador y plano coinciden en el punto $D$.

Igualmente en la Fig.\ref{RotaT} se puede observar también que el ángulo $\beta$ adicional que barre el observador está relacionado con los ángulos $\gamma$ y $\alpha_{f}$ mediante 
\begin{equation}\label{R2}
\beta+\gamma+\alpha_{f}=\pi.
\end{equation}
Además, de los triángulos $\widehat{qAB}$ y $\widehat{qCD}$ se encuentran las relaciones 
\begin{equation}\label{R3}
2\alpha_{i}+\theta_{i}=\pi.
\end{equation}
\begin{equation}\label{R4}
2\alpha_{f}+\theta_{f}=\pi.
\end{equation}
Ahora, al combinar las ecuaciones  (\ref{R1}),(\ref{R2}),(\ref{R3}) y (\ref{R4}) se encuentra 
\begin{equation}\label{R5}
\beta=\frac{ \theta_{f}-\theta_{i} }{2}+\Omega\tau,
\end{equation}
donde $\theta_{i}=\theta(t)$ y $\theta_{f}=\theta(t+\tau)$ con
\begin{equation}\label{R6}
 \theta(t)=2\cos^{-1}\left(\tan\phi \: \tan\epsilon \cos\left(\Omega t\right) \right ).
\end{equation}
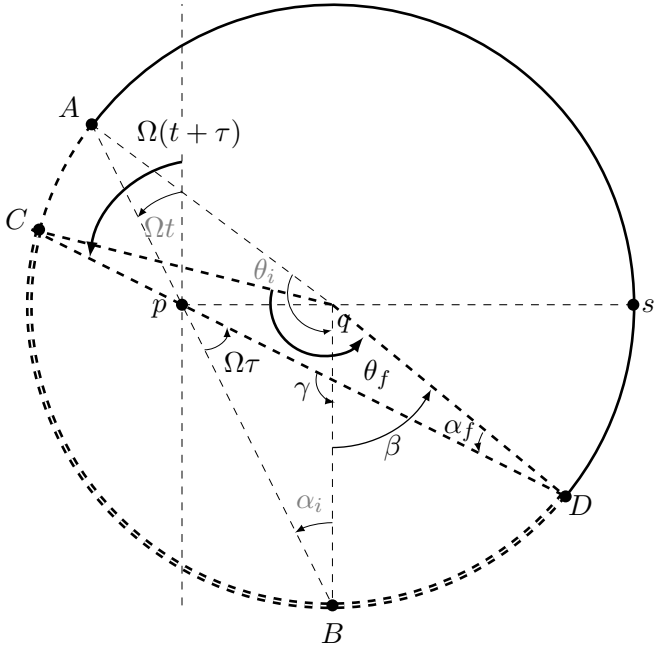
\begin{figure}[ht]
\begin{center}
\begin{tikzpicture}[xscale=1.0,yscale=1.0]
\draw[line width=1pt] (3.1cm,-2.55cm) arc (320.56:502.5:4cm);
\draw[dashed, line width=1pt] (-3.2cm,2.4cm) arc (502.5:525.61:4cm);
\draw[dashed,double,line width=1pt] (-3.9cm,1.0cm) arc (165.44:320.56:4cm);
\draw[dashed,black,line width=0.3pt] (-3.2cm,2.4cm)--(0cm,-4cm);
\draw[dashed,black,line width=0.3pt] (-3.2cm,2.4cm)--(0cm,0cm);
\draw[dashed,black,line width=1pt] (-4cm,1cm)--(0cm,0cm);
\draw[dashed,black,line width=0.3pt] (0cm,0cm)--(0cm,-4cm);
\draw[dashed,black,line width=1pt] (0cm,0cm)--(3.1cm,-2.55cm);
\draw[dashed,black] (-2cm,0cm)--(4cm,0cm);
\draw[dashed,black,line width=0.3pt] (-2cm,-4cm)--(-2cm,2.0cm);
\draw[dashed,black,line width=0.3pt] (-2cm,2.5cm)--(-2cm,4cm);
\draw[dashed,black,line width=1pt] (-4cm,1cm)--(3.1cm,-2.55cm);
\fill[black] (-3.2cm,2.4cm) circle(0.08cm); 
\fill[black] (0cm,-4cm) circle(0.08cm); 
\fill[black] (-3.9cm,1.0cm) circle(0.08cm); 
\fill[black] (3.1cm,-2.55cm) circle(0.08cm); 
\fill[black] (4cm,0cm) circle(0.08cm); 
\fill[black] (-2cm,0cm) circle(0.08cm); 
\draw[-latex,line width=0.1pt] (-0.53cm,0.4cm) arc (150:280:0.5cm);
\draw[-latex,line width=1pt] (-0.8cm,0.2cm) arc (165:320:0.7cm);
\draw[-latex,line width=0.3pt] (-2.0cm,1.5cm) arc (100:140:1.0cm);
\draw[-latex,line width=1pt] (-2.0cm,1.9cm) arc (100:172:1.5cm);
\draw[-latex,line width=0.3pt] (0cm,-2.9cm) arc (90:120:1cm);
\draw[-latex,line width=0.3pt] (2cm,-1.7cm) arc (150:190:0.4cm);
\draw[-latex,line width=0.3pt] (-0.2cm,-0.9cm) arc (160:261:0.3cm);
\draw[-latex,line width=0.3pt] (-1.7cm,-0.6cm) arc (270:360:0.3cm);
\draw[-latex,line width=0.5pt] (0cm,-1.9cm) arc (270:331:1.5cm);
\coordinate [label=below:\textcolor{black} {$q$}] (x) at  (0.15cm,0.0cm);
\coordinate [label=below:\textcolor{black} {$A$}] (x) at  (-3.5cm,2.9cm);
\coordinate [label=below:\textcolor{black} {$B$}] (x) at  (0cm,-4.1cm);
\coordinate [label=below:\textcolor{black} {$s$}] (x) at  (4.2cm,0.2cm);
\coordinate [label=below:\textcolor{black} {$C$}] (x) at  (-4.2cm,1.4cm);
\coordinate [label=below:\textcolor{black} {$D$}] (x) at  (3.3cm,-2.4cm);
\coordinate [label=below:\textcolor{black} {$p$}] (x) at  (-2.3cm,0.2cm);
\coordinate [label=below:\textcolor{gray} {$\theta_{i}$}] (x) at  (-0.9cm,0.75cm);
\coordinate [label=below:\textcolor{black} {$\theta_{f}$}] (x) at  (0.6cm,-0.6cm);
\coordinate [label=below:\textcolor{gray} {$\alpha _{i}$}] (x) at  (-0.3cm,-2.4cm);
\coordinate [label=below:\textcolor{black} {$\alpha _{f}$}] (x) at  (1.7cm,-1.4cm);
\coordinate [label=below:\textcolor{black} {$\beta$}] (x) at  (0.8cm,-1.6cm);
\coordinate [label=below:\textcolor{gray} {$\Omega t$}] (x) at  (-2.3cm,1.3cm);
\coordinate [label=below:\textcolor{black} {$\Omega \tau$}] (x) at  (-1.2cm,-0.5cm);
\coordinate [label=below:\textcolor{black} {$\gamma$}] (x) at  (-0.4cm,-0.9cm);
\coordinate [label=below:\textcolor{black} {$\Omega (t+ \tau)$}] (x) at  (-1.9cm,2.6cm);
\end{tikzpicture}
\caption{Zonas iluminadas al principio y al final de una fracci\'on de revoluci\'on terrestre.}
\label{RotaT}
\end{center}
\end{figure}
Cuando la Tierra gira alrededor del Sol la zona iluminada cambia de $\theta_{i}$ a 
$\theta_{f}$ lo que implica que un observador que coincidida con el plano de luz inicialmente en $A$ barre un arco $\stackrel{\textstyle\frown}{\mathrm{AD}}$ y emplea un tiempo $\tau$ que se concreta justo cuando el observador alcanza el plano de luz en $D$. De manera que  
\begin{equation}\label{TIME1}
 \theta_{i}+\beta=\omega \tau,
\end{equation}
siendo $\tau$ el tiempo que el observador está iluminado. Ahora, como  $t_{D}$ el tiempo que el observador está iluminado cuando no se toma en cuenta la pequeña traslación que la Tierra sufre mientras hace una revolución sobre su eje $\Delta t=\tau-\tau_{i}$ es el tiempo adicional de iluminación que debe soportar el observador si se toma en cuenta esta pequeña traslación. 

Para calcular $\tau$ se reemplaza  la ecuación (\ref{R5}) en (\ref{TIME1}) y se  encuentra    
\begin{equation}\label{TIME2}
 \tau=\frac{\theta_{f}+\theta_{i}}{2(\omega-\Omega)}.
\end{equation}
En (\ref{TIME2}) se espera que $\theta_{f}\approx \theta_{i}$ dado que $\Omega<<\omega$, lo que sugiere hacer una expansión en serie de Taylor de $\theta(\tau)$ alrededor de $0$ y tomar la aproximación con los primeros dos términos de la serie, así se obtiene 
\begin{equation}\label{TIME3}
 \theta_{f}=\theta_{i}+2\eta \tau
\end{equation}

con 
\begin{equation}\label{TIME4A}
 \eta=\frac{\Omega \tan \epsilon\: \tan \phi \:\sin(\Omega t) }{\sqrt{1-\tan^2\epsilon \tan^2\phi \cos^2(\Omega t)}}, 
\end{equation}
que también cumple la condición (\ref{cond}) y los términos de coseno y seno junto con los ordenes de magnitud garantizan que (\ref{TIME4A}) siempre permanece finito y menor que $1\frac{rad}{s}$.

Para encontrar la corrección por traslación $\Delta t$ se reemplaza la ecuación (\ref{TIME3}) en (\ref{TIME2}), obteniendo
\begin{equation}\label{TIME4}
 \tau=\frac{\theta_{i}}{\omega-\Omega-\eta}
\end{equation}
lo que conduce a
\begin{equation}\label{TIME5}
 \Delta t=\frac{\left(\Omega-\eta \right)\theta_{i}}{\omega \left(\omega-\Omega-\eta\right)},
\end{equation}
que es la correción por traslación buscada y resulta ser de $\approx2^{m}$.

De otro lado, se sabe que cuando los rayos de luz penetran en la atmósfera sufren refracción que aumenta el tiempo de iluminación, así en la Fig.\ref{SolsPNT} los rayos de luz paralelos a la semirecta $\overline{ps}$ que llegan tanto al punto $A$ como al $B$ convergen un poco hacia el punto $q$. De esta manera un observador comienza a ser iluminado un poco antes de llegar al punto $A$ y sigue un poco iluminado después de pasar por el punto $B$. En la literatura suele tomarse  la porción de ángulo que barre un observador bajo los efectos de la luz refractada como $\theta_R \approx 9,95\times 10^{-3}rad$ \cite{Portilla, Aus}. De lo anterior se entiende que este fenómeno genera un tiempo adicional de iluminación dado por $\frac{2\theta_R}{\omega}$, que es aproximadamente $4^{m}5^{s}$.

De acuerdo con las anteriores consideraciones la ecuación (\ref{time2}) toman la forma
\begin{equation}\label{Cor1}
t_{D}=\Delta t+\frac{T}{\pi}\left[ 2\theta_R+\cos^{-1}\left(\tan\phi \: \tan\epsilon \cos\left(\Omega t\right) \right ) \right].
\end{equation}
\section{Modelo vs resultados numéricos}
De los datos numéricos ofrecidos por la simulación de la Geoscience Australia Agency  puede obtenerse resultados que calculan el tiempo que un observador permanece iluminado durante una revolución de la Tierra. Los aspectos más relevantes para la confrontación del modelo analítico que toma en cuenta el modelo númerico de Geoscience Australia Agency  son la trayectoria elíptica  de la Tierra alrededor del Sol y la refracción estándar de la luz debida a la atmósfera \cite{Aus}.     
\begin {figure}
\begin{center}
\input{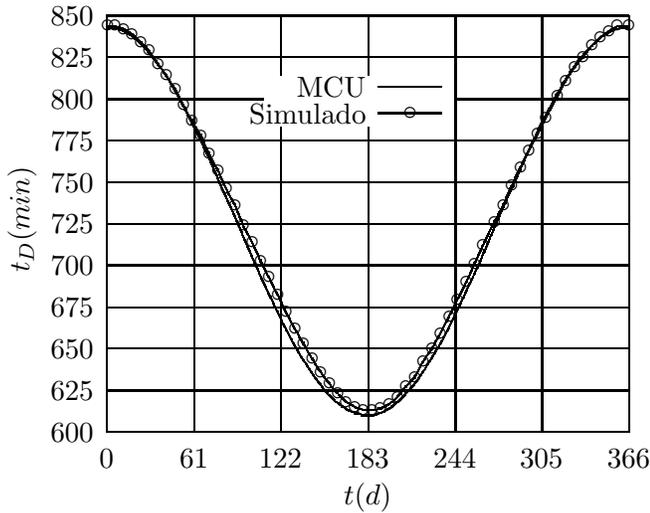}
\caption{Modelo analítico con MCU  vs simulación para $\phi=30^{o}$ y $\epsilon=23.5^{o}$ entre el $21$ de junio de $2009$ y el $21$ de junio de $2010$.}
\label{tD_N}
\end{center}
\end {figure}

En la Fig. (\ref{tD_N}) se muestra el tiempo de iluminación $t_D$ en minutos como función del tiempo  $t$  transcurrido en días desde el comienzo del solsticio de verano de $2009$ hasta justo el comienzo  del solsticio de verano de $2010$, se observa  mayor diferencia entre los tiempos obtenidos con la simulación y el modelo analítico  desde mediados de febrero hasta los primeros días de octubre época del año en la cual la Tierra está ceca al perihelio, que ocurre aproximadamente el 4 de enero\cite{Nasa1}. Por el contrario  se observa un buen acuerdo entre mediados de octubre y los primeros días de febrero época del año en que la Tierra está cerca a su afelio, que ocurre aproximadamente el 4  de julio\cite{Nasa2}.  La tercera ley de Kepler implica que cuando la Tierra se encuentra en el perihelio su rapidez es mayor que cuando se encuentra en el afelio lo que trae como consecuencia que el tiempo adicional de iluminación por incremento de traslación aumente  ya que el plano $M$ de la Fig.\ref{SolsI} se mueve más rápido sobre el punto $q$ y por tanto un observador sobre la Tierra tarda más tiempo en alcanzar el plano de luz solar. Por el contrario cuando la Tierra se encuentra en el afelio el plano $M$  se mueve más lento produciendo que el tiempo adicional de iluminación por traslación sea menor. Consecuentemente en cecanias del perihelio la simulación muestra resultados alejados del modelo analítico de trayectoria circular  que corresponde a unos $\approx12.5^{m}$ que trae consigo  un error relativo porcentual de $\approx1.6\% $. 

\begin {figure}
\begin{center}
\input{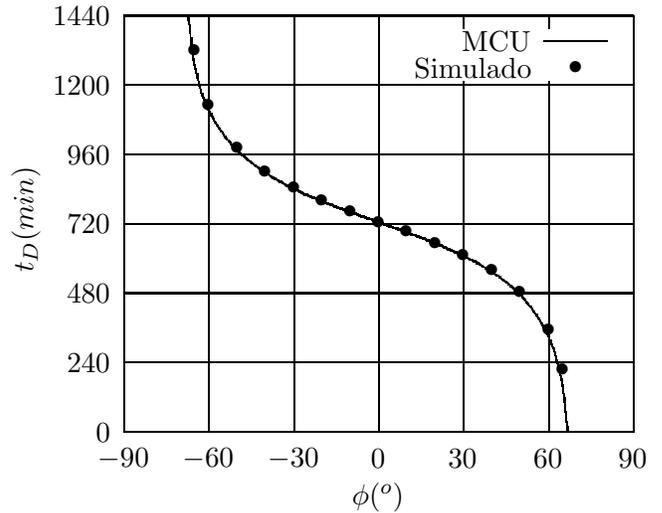}
\caption{Modelo analítico con MCU vs modelo numérico para el solsticio de invierno en el hemisferio norte.}
\label{t_sol}
\end{center}
\end {figure}

De otro lado la Fig. (\ref{t_sol}) muestra en el solsticio de invierno el tiempo de iluminación para diferentes latitudes, se observa un muy buen acuerdo entre el modelo simulado y el modelo analítico.   

Diversas comparaciones entre el modelo analítico y la simulación arrojan resultados  similares. Así, el modelo analítico da cuenta de manera cercana y con un mecanismo físico de explicación al tiempo de iluminación solar en cualquier tiempo del año y en cualquier lugar de la Tierra.
\section{Algunas sugerencias didácticas}   
De acuerdo con los resultados obtenidos la ecuación (\ref{Cor1}) permite estimar el tiempo  de iluminación para una fecha particular del año y para un lugar particular en la Tierra caracterizado por su latitud, tomando  correcciones por refracción y traslación. Con este resultado  es posible atacar problemas de la vida cotidiana tales como  la estimación del tiempo de iluminación solar en un país particular;  apoyar una explicación física cuantitativa de las estaciones del año; ofrecer un espacio para conceptualizar el MCU ejemplificando a través de los movimientos de traslación y rotación de la Tierra; usar esquemas como el mostrado en la Fig.\ref{Rotada} para plantear una discusión sobre la refracción de la luz por la atmósfera; discutir los fenómenos del Sol de media noche y la noche polar, en tre otros.  Así mismo, se pueden plantear situaciones en las cuales dada una fecha particular del año se pregunta por la duración de la iluminación en diferentes países y ofrecer explicación física de las diferencias presentadas.     
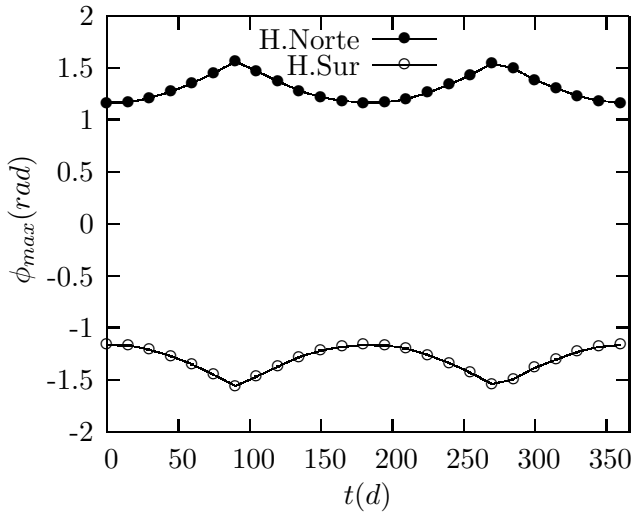
\begin {figure}
\begin{center}
\setlength{\unitlength}{0.240900pt}
\ifx\plotpoint\undefined\newsavebox{\plotpoint}\fi
\sbox{\plotpoint}{\rule[-0.200pt]{0.400pt}{0.400pt}}%
\begin{picture}(1062,826)(0,0)
\sbox{\plotpoint}{\rule[-0.200pt]{0.400pt}{0.400pt}}%
\put(191.0,131.0){\rule[-0.200pt]{4.818pt}{0.400pt}}
\put(171,131){\makebox(0,0)[r]{-2}}
\put(991.0,131.0){\rule[-0.200pt]{4.818pt}{0.400pt}}
\put(191.0,213.0){\rule[-0.200pt]{4.818pt}{0.400pt}}
\put(171,213){\makebox(0,0)[r]{-1.5}}
\put(991.0,213.0){\rule[-0.200pt]{4.818pt}{0.400pt}}
\put(191.0,295.0){\rule[-0.200pt]{4.818pt}{0.400pt}}
\put(171,295){\makebox(0,0)[r]{-1}}
\put(991.0,295.0){\rule[-0.200pt]{4.818pt}{0.400pt}}
\put(191.0,376.0){\rule[-0.200pt]{4.818pt}{0.400pt}}
\put(171,376){\makebox(0,0)[r]{-0.5}}
\put(991.0,376.0){\rule[-0.200pt]{4.818pt}{0.400pt}}
\put(191.0,458.0){\rule[-0.200pt]{4.818pt}{0.400pt}}
\put(171,458){\makebox(0,0)[r]{ 0}}
\put(991.0,458.0){\rule[-0.200pt]{4.818pt}{0.400pt}}
\put(191.0,540.0){\rule[-0.200pt]{4.818pt}{0.400pt}}
\put(171,540){\makebox(0,0)[r]{ 0.5}}
\put(991.0,540.0){\rule[-0.200pt]{4.818pt}{0.400pt}}
\put(191.0,622.0){\rule[-0.200pt]{4.818pt}{0.400pt}}
\put(171,622){\makebox(0,0)[r]{ 1}}
\put(991.0,622.0){\rule[-0.200pt]{4.818pt}{0.400pt}}
\put(191.0,703.0){\rule[-0.200pt]{4.818pt}{0.400pt}}
\put(171,703){\makebox(0,0)[r]{ 1.5}}
\put(991.0,703.0){\rule[-0.200pt]{4.818pt}{0.400pt}}
\put(191.0,785.0){\rule[-0.200pt]{4.818pt}{0.400pt}}
\put(171,785){\makebox(0,0)[r]{ 2}}
\put(991.0,785.0){\rule[-0.200pt]{4.818pt}{0.400pt}}
\put(191.0,131.0){\rule[-0.200pt]{0.400pt}{4.818pt}}
\put(191,90){\makebox(0,0){ 0}}
\put(191.0,765.0){\rule[-0.200pt]{0.400pt}{4.818pt}}
\put(303.0,131.0){\rule[-0.200pt]{0.400pt}{4.818pt}}
\put(303,90){\makebox(0,0){ 50}}
\put(303.0,765.0){\rule[-0.200pt]{0.400pt}{4.818pt}}
\put(415.0,131.0){\rule[-0.200pt]{0.400pt}{4.818pt}}
\put(415,90){\makebox(0,0){ 100}}
\put(415.0,765.0){\rule[-0.200pt]{0.400pt}{4.818pt}}
\put(527.0,131.0){\rule[-0.200pt]{0.400pt}{4.818pt}}
\put(527,90){\makebox(0,0){ 150}}
\put(527.0,765.0){\rule[-0.200pt]{0.400pt}{4.818pt}}
\put(639.0,131.0){\rule[-0.200pt]{0.400pt}{4.818pt}}
\put(639,90){\makebox(0,0){ 200}}
\put(639.0,765.0){\rule[-0.200pt]{0.400pt}{4.818pt}}
\put(751.0,131.0){\rule[-0.200pt]{0.400pt}{4.818pt}}
\put(751,90){\makebox(0,0){ 250}}
\put(751.0,765.0){\rule[-0.200pt]{0.400pt}{4.818pt}}
\put(863.0,131.0){\rule[-0.200pt]{0.400pt}{4.818pt}}
\put(863,90){\makebox(0,0){ 300}}
\put(863.0,765.0){\rule[-0.200pt]{0.400pt}{4.818pt}}
\put(975.0,131.0){\rule[-0.200pt]{0.400pt}{4.818pt}}
\put(975,90){\makebox(0,0){ 350}}
\put(975.0,765.0){\rule[-0.200pt]{0.400pt}{4.818pt}}
\put(191.0,131.0){\rule[-0.200pt]{0.400pt}{157.549pt}}
\put(191.0,131.0){\rule[-0.200pt]{197.538pt}{0.400pt}}
\put(1011.0,131.0){\rule[-0.200pt]{0.400pt}{157.549pt}}
\put(191.0,785.0){\rule[-0.200pt]{197.538pt}{0.400pt}}
\put(50,458){\makebox(0,0){$\rotatebox{90} { \parbox{4cm}{\begin{align*} \phi_{max}(rad) \end{align*}}}$}}
\put(601,29){\makebox(0,0){$t(d)$}}
\put(589,749){\makebox(0,0)[r]{H.Norte}}
\put(609.0,749.0){\rule[-0.200pt]{24.090pt}{0.400pt}}
\put(191,648){\usebox{\plotpoint}}
\put(191,647.67){\rule{8.191pt}{0.400pt}}
\multiput(191.00,647.17)(17.000,1.000){2}{\rule{4.095pt}{0.400pt}}
\multiput(225.00,649.59)(2.476,0.485){11}{\rule{1.986pt}{0.117pt}}
\multiput(225.00,648.17)(28.879,7.000){2}{\rule{0.993pt}{0.400pt}}
\multiput(258.00,656.58)(1.746,0.491){17}{\rule{1.460pt}{0.118pt}}
\multiput(258.00,655.17)(30.970,10.000){2}{\rule{0.730pt}{0.400pt}}
\multiput(292.00,666.58)(1.290,0.493){23}{\rule{1.115pt}{0.119pt}}
\multiput(292.00,665.17)(30.685,13.000){2}{\rule{0.558pt}{0.400pt}}
\multiput(325.00,679.58)(1.073,0.494){29}{\rule{0.950pt}{0.119pt}}
\multiput(325.00,678.17)(32.028,16.000){2}{\rule{0.475pt}{0.400pt}}
\multiput(359.00,695.58)(0.952,0.495){33}{\rule{0.856pt}{0.119pt}}
\multiput(359.00,694.17)(32.224,18.000){2}{\rule{0.428pt}{0.400pt}}
\multiput(393.00,711.92)(1.113,-0.494){27}{\rule{0.980pt}{0.119pt}}
\multiput(393.00,712.17)(30.966,-15.000){2}{\rule{0.490pt}{0.400pt}}
\multiput(426.00,696.92)(1.073,-0.494){29}{\rule{0.950pt}{0.119pt}}
\multiput(426.00,697.17)(32.028,-16.000){2}{\rule{0.475pt}{0.400pt}}
\multiput(460.00,680.92)(1.113,-0.494){27}{\rule{0.980pt}{0.119pt}}
\multiput(460.00,681.17)(30.966,-15.000){2}{\rule{0.490pt}{0.400pt}}
\multiput(493.00,665.92)(1.746,-0.491){17}{\rule{1.460pt}{0.118pt}}
\multiput(493.00,666.17)(30.970,-10.000){2}{\rule{0.730pt}{0.400pt}}
\multiput(527.00,655.93)(3.022,-0.482){9}{\rule{2.367pt}{0.116pt}}
\multiput(527.00,656.17)(29.088,-6.000){2}{\rule{1.183pt}{0.400pt}}
\multiput(561.00,649.95)(7.160,-0.447){3}{\rule{4.500pt}{0.108pt}}
\multiput(561.00,650.17)(23.660,-3.000){2}{\rule{2.250pt}{0.400pt}}
\put(594,647.67){\rule{8.191pt}{0.400pt}}
\multiput(594.00,647.17)(17.000,1.000){2}{\rule{4.095pt}{0.400pt}}
\multiput(628.00,649.59)(3.604,0.477){7}{\rule{2.740pt}{0.115pt}}
\multiput(628.00,648.17)(27.313,5.000){2}{\rule{1.370pt}{0.400pt}}
\multiput(661.00,654.58)(1.746,0.491){17}{\rule{1.460pt}{0.118pt}}
\multiput(661.00,653.17)(30.970,10.000){2}{\rule{0.730pt}{0.400pt}}
\multiput(695.00,664.58)(1.329,0.493){23}{\rule{1.146pt}{0.119pt}}
\multiput(695.00,663.17)(31.621,13.000){2}{\rule{0.573pt}{0.400pt}}
\multiput(729.00,677.58)(1.113,0.494){27}{\rule{0.980pt}{0.119pt}}
\multiput(729.00,676.17)(30.966,15.000){2}{\rule{0.490pt}{0.400pt}}
\multiput(762.00,692.58)(0.952,0.495){33}{\rule{0.856pt}{0.119pt}}
\multiput(762.00,691.17)(32.224,18.000){2}{\rule{0.428pt}{0.400pt}}
\multiput(796.00,708.93)(2.211,-0.488){13}{\rule{1.800pt}{0.117pt}}
\multiput(796.00,709.17)(30.264,-8.000){2}{\rule{0.900pt}{0.400pt}}
\multiput(830.00,700.92)(0.923,-0.495){33}{\rule{0.833pt}{0.119pt}}
\multiput(830.00,701.17)(31.270,-18.000){2}{\rule{0.417pt}{0.400pt}}
\multiput(863.00,682.92)(1.329,-0.493){23}{\rule{1.146pt}{0.119pt}}
\multiput(863.00,683.17)(31.621,-13.000){2}{\rule{0.573pt}{0.400pt}}
\multiput(897.00,669.92)(1.401,-0.492){21}{\rule{1.200pt}{0.119pt}}
\multiput(897.00,670.17)(30.509,-12.000){2}{\rule{0.600pt}{0.400pt}}
\multiput(930.00,657.93)(2.211,-0.488){13}{\rule{1.800pt}{0.117pt}}
\multiput(930.00,658.17)(30.264,-8.000){2}{\rule{0.900pt}{0.400pt}}
\multiput(964.00,649.95)(7.383,-0.447){3}{\rule{4.633pt}{0.108pt}}
\multiput(964.00,650.17)(24.383,-3.000){2}{\rule{2.317pt}{0.400pt}}
\put(191,648){\makebox(0,0){$\bullet$}}
\put(225,649){\makebox(0,0){$\bullet$}}
\put(258,656){\makebox(0,0){$\bullet$}}
\put(292,666){\makebox(0,0){$\bullet$}}
\put(325,679){\makebox(0,0){$\bullet$}}
\put(359,695){\makebox(0,0){$\bullet$}}
\put(393,713){\makebox(0,0){$\bullet$}}
\put(426,698){\makebox(0,0){$\bullet$}}
\put(460,682){\makebox(0,0){$\bullet$}}
\put(493,667){\makebox(0,0){$\bullet$}}
\put(527,657){\makebox(0,0){$\bullet$}}
\put(561,651){\makebox(0,0){$\bullet$}}
\put(594,648){\makebox(0,0){$\bullet$}}
\put(628,649){\makebox(0,0){$\bullet$}}
\put(661,654){\makebox(0,0){$\bullet$}}
\put(695,664){\makebox(0,0){$\bullet$}}
\put(729,677){\makebox(0,0){$\bullet$}}
\put(762,692){\makebox(0,0){$\bullet$}}
\put(796,710){\makebox(0,0){$\bullet$}}
\put(830,702){\makebox(0,0){$\bullet$}}
\put(863,684){\makebox(0,0){$\bullet$}}
\put(897,671){\makebox(0,0){$\bullet$}}
\put(930,659){\makebox(0,0){$\bullet$}}
\put(964,651){\makebox(0,0){$\bullet$}}
\put(998,648){\makebox(0,0){$\bullet$}}
\put(659,749){\makebox(0,0){$\bullet$}}
\put(589,708){\makebox(0,0)[r]{H.Sur}}
\put(609.0,708.0){\rule[-0.200pt]{24.090pt}{0.400pt}}
\put(191,268){\usebox{\plotpoint}}
\put(191,266.67){\rule{8.191pt}{0.400pt}}
\multiput(191.00,267.17)(17.000,-1.000){2}{\rule{4.095pt}{0.400pt}}
\multiput(225.00,265.93)(2.476,-0.485){11}{\rule{1.986pt}{0.117pt}}
\multiput(225.00,266.17)(28.879,-7.000){2}{\rule{0.993pt}{0.400pt}}
\multiput(258.00,258.92)(1.746,-0.491){17}{\rule{1.460pt}{0.118pt}}
\multiput(258.00,259.17)(30.970,-10.000){2}{\rule{0.730pt}{0.400pt}}
\multiput(292.00,248.92)(1.290,-0.493){23}{\rule{1.115pt}{0.119pt}}
\multiput(292.00,249.17)(30.685,-13.000){2}{\rule{0.558pt}{0.400pt}}
\multiput(325.00,235.92)(1.073,-0.494){29}{\rule{0.950pt}{0.119pt}}
\multiput(325.00,236.17)(32.028,-16.000){2}{\rule{0.475pt}{0.400pt}}
\multiput(359.00,219.92)(0.952,-0.495){33}{\rule{0.856pt}{0.119pt}}
\multiput(359.00,220.17)(32.224,-18.000){2}{\rule{0.428pt}{0.400pt}}
\multiput(393.00,203.58)(1.113,0.494){27}{\rule{0.980pt}{0.119pt}}
\multiput(393.00,202.17)(30.966,15.000){2}{\rule{0.490pt}{0.400pt}}
\multiput(426.00,218.58)(1.073,0.494){29}{\rule{0.950pt}{0.119pt}}
\multiput(426.00,217.17)(32.028,16.000){2}{\rule{0.475pt}{0.400pt}}
\multiput(460.00,234.58)(1.113,0.494){27}{\rule{0.980pt}{0.119pt}}
\multiput(460.00,233.17)(30.966,15.000){2}{\rule{0.490pt}{0.400pt}}
\multiput(493.00,249.58)(1.746,0.491){17}{\rule{1.460pt}{0.118pt}}
\multiput(493.00,248.17)(30.970,10.000){2}{\rule{0.730pt}{0.400pt}}
\multiput(527.00,259.59)(3.022,0.482){9}{\rule{2.367pt}{0.116pt}}
\multiput(527.00,258.17)(29.088,6.000){2}{\rule{1.183pt}{0.400pt}}
\multiput(561.00,265.61)(7.160,0.447){3}{\rule{4.500pt}{0.108pt}}
\multiput(561.00,264.17)(23.660,3.000){2}{\rule{2.250pt}{0.400pt}}
\put(594,266.67){\rule{8.191pt}{0.400pt}}
\multiput(594.00,267.17)(17.000,-1.000){2}{\rule{4.095pt}{0.400pt}}
\multiput(628.00,265.93)(3.604,-0.477){7}{\rule{2.740pt}{0.115pt}}
\multiput(628.00,266.17)(27.313,-5.000){2}{\rule{1.370pt}{0.400pt}}
\multiput(661.00,260.92)(1.746,-0.491){17}{\rule{1.460pt}{0.118pt}}
\multiput(661.00,261.17)(30.970,-10.000){2}{\rule{0.730pt}{0.400pt}}
\multiput(695.00,250.92)(1.329,-0.493){23}{\rule{1.146pt}{0.119pt}}
\multiput(695.00,251.17)(31.621,-13.000){2}{\rule{0.573pt}{0.400pt}}
\multiput(729.00,237.92)(1.113,-0.494){27}{\rule{0.980pt}{0.119pt}}
\multiput(729.00,238.17)(30.966,-15.000){2}{\rule{0.490pt}{0.400pt}}
\multiput(762.00,222.92)(0.952,-0.495){33}{\rule{0.856pt}{0.119pt}}
\multiput(762.00,223.17)(32.224,-18.000){2}{\rule{0.428pt}{0.400pt}}
\multiput(796.00,206.59)(2.211,0.488){13}{\rule{1.800pt}{0.117pt}}
\multiput(796.00,205.17)(30.264,8.000){2}{\rule{0.900pt}{0.400pt}}
\multiput(830.00,214.58)(0.923,0.495){33}{\rule{0.833pt}{0.119pt}}
\multiput(830.00,213.17)(31.270,18.000){2}{\rule{0.417pt}{0.400pt}}
\multiput(863.00,232.58)(1.329,0.493){23}{\rule{1.146pt}{0.119pt}}
\multiput(863.00,231.17)(31.621,13.000){2}{\rule{0.573pt}{0.400pt}}
\multiput(897.00,245.58)(1.401,0.492){21}{\rule{1.200pt}{0.119pt}}
\multiput(897.00,244.17)(30.509,12.000){2}{\rule{0.600pt}{0.400pt}}
\multiput(930.00,257.59)(2.211,0.488){13}{\rule{1.800pt}{0.117pt}}
\multiput(930.00,256.17)(30.264,8.000){2}{\rule{0.900pt}{0.400pt}}
\multiput(964.00,265.61)(7.383,0.447){3}{\rule{4.633pt}{0.108pt}}
\multiput(964.00,264.17)(24.383,3.000){2}{\rule{2.317pt}{0.400pt}}
\put(191,268){\makebox(0,0){$\circ$}}
\put(225,267){\makebox(0,0){$\circ$}}
\put(258,260){\makebox(0,0){$\circ$}}
\put(292,250){\makebox(0,0){$\circ$}}
\put(325,237){\makebox(0,0){$\circ$}}
\put(359,221){\makebox(0,0){$\circ$}}
\put(393,203){\makebox(0,0){$\circ$}}
\put(426,218){\makebox(0,0){$\circ$}}
\put(460,234){\makebox(0,0){$\circ$}}
\put(493,249){\makebox(0,0){$\circ$}}
\put(527,259){\makebox(0,0){$\circ$}}
\put(561,265){\makebox(0,0){$\circ$}}
\put(594,268){\makebox(0,0){$\circ$}}
\put(628,267){\makebox(0,0){$\circ$}}
\put(661,262){\makebox(0,0){$\circ$}}
\put(695,252){\makebox(0,0){$\circ$}}
\put(729,239){\makebox(0,0){$\circ$}}
\put(762,224){\makebox(0,0){$\circ$}}
\put(796,206){\makebox(0,0){$\circ$}}
\put(830,214){\makebox(0,0){$\circ$}}
\put(863,232){\makebox(0,0){$\circ$}}
\put(897,245){\makebox(0,0){$\circ$}}
\put(930,257){\makebox(0,0){$\circ$}}
\put(964,265){\makebox(0,0){$\circ$}}
\put(998,268){\makebox(0,0){$\circ$}}
\put(659,708){\makebox(0,0){$\circ$}}
\put(191.0,131.0){\rule[-0.200pt]{0.400pt}{157.549pt}}
\put(191.0,131.0){\rule[-0.200pt]{197.538pt}{0.400pt}}
\put(1011.0,131.0){\rule[-0.200pt]{0.400pt}{157.549pt}}
\put(191.0,785.0){\rule[-0.200pt]{197.538pt}{0.400pt}}
\end{picture}

\caption{Latitud para el Sol de media noche y la noche polar con el transcurso del año sin correciones por refracción ni traslación.}
\label{noches}
\end{center}
\end {figure}
Resulta particularmente interesante usar el modelo para estimar la fecha de inicio de los días y noches polares, para esto es necesario realizar análisis gráfico de la ecuación (\ref{Lmax}) que muestra la máxima latitud $\phi_{max}$ que es posible evaluar en las ecuaciones   (\ref{time1}) y (\ref{Cor1}) como función del tiempo y que a su vez permite estudiar las zonas de la Tierra donde comienza el sol de media noche o la noche polar. 

En la  Fig.\ref{noches} se muestra la evolución de la latitud máxima con el transcurso del año para los hemisferios norte y sur, así  por arriba de la línea de puntos oscuros habrá latitudes que están en noche polar para una fecha determinada del año y por debajo de la línea de puntos claros habrá latitudes en las cuales  habrá día polar.

Además, el modelo puede usarse para estudiar algunos aspectos referidos a la iluminación de planetas tonto en el sistema solar como en otros sistemas solares.
\section{Conclusiones}
Se demostró que es posible formular un modelo analítico con base en la hipotesis de un MCU de la Tierra alrededor del Sol que permite estimar el tiempo de iluminación para un tiempo cualquiera después del solsticio de invierno en el hemisferio norte a cualquier latitud, que concuerda  bien con el modelo numérico  de la Geoscience Australia Agency y se realizaron correcciones por traslación y refracción siendo respectivamente de $\approx2^m$ y de $\approx4^{m}5^{s} $, el modelo brinda un escenario rico en fenómenos terrestres relacionados con la astronomía   que permiten estudiar algunos aspectos de la física básica. Queda como problema sugerido aplicar el modelo a planetas extrasolares con diversas características para estudiar diferentes aspectos relacionados con el tiempo de iluminación sobre sus superficies. 
\section*{Agradecimientos}
Los autores agradecen a los profesores Fabian Galindo, Giovanni Cardona y Guillermo Avendaño por la lectura y las sugerencias realizadas sobre el manuscrito;  al Departamento de Ciencias Naturales de la Universidad  Central por el apoyo y el tiempo asignado a la investigación y al CICATA del IPN de México por su continua colaboración.
\renewcommand{\refname}{Referencias}

\end{document}